\documentclass[11pt]{article}             
\usepackage{osid}                         %

\usepackage{graphicx}
\usepackage{amsmath}
\usepackage{epsfig}
\usepackage{dcolumn}
\usepackage{bm}
\usepackage{color}
\usepackage{epstopdf}
\usepackage{enumerate}

\newcommand{\nc}{\newcommand}
\nc{\eq}{\begin{equation}}
\nc{\eeq}{\end{equation}}
\nc{\eqa}{\begin{eqnarray}}
\nc{\eeqa}{\end{eqnarray}}
\nc{\ar}{\begin{array}}
\nc{\ear}{\end{array}}
\nc{\bfig}{\begin{figure}}
\nc{\efig}{\end{figure}}
\nc{\dg}{\dagger}
\nc{\sx}{\sigma_x}
\nc{\sy}{\sigma_y}
\nc{\sz}{\sigma_z}
\nc{\spl}{\sigma_+}
\nc{\sm}{\sigma_-}
\nc{\nn}{\nonumber}
\nc{\noi}{\noindent}
\nc{\adg}{a^{\dg}}
\nc{\kvec}{\mathbf{k}}
\nc{\xvec}{\mathbf{x}}

\def\bra#1{\mathinner{\langle{#1}|}}
\def\ket#1{\mathinner{|{#1}\rangle}}

\title{Non-Markovian quantum probes}
\author{P.~Haikka\\{\footnotesize\it Turku Center for Quantum Physics, Department of
Physics and Astronomy, University of Turku, FIN20014, Turku,
Finland \\ pmehai@utu.fil}\\[2ex]
         S.~Maniscalco\\{\footnotesize\it Turku Center for Quantum Physics, Department of
Physics and Astronomy, University of Turku, FIN20014, Turku,
Finland \\EPS/Physics, Heriot-Watt University, Edinburgh, EH14 4AS, United Kingdom \\ s.maniscalco@hw.ac.uk} }

\begin{document}

\maketitle
\begin{abstract}
We review the most recent developments in the theory of open quantum systems focusing on situations in which the reservoir memory effects, due to long-lasting and non-negligible correlations between system and environment, play a crucial role. These systems are often referred to as non-Markovian systems. After a brief summary of different measures of non-Markovianity that have been introduced over the last few years we restrict our analysis to the investigation of information flow between system and environment. Within this framework we introduce an important application of non-Markovianity, namely its use as a quantum probe of complex quantum systems. To illustrate this point we consider quantum probes of ultracold gases, spin chains, and trapped ion crystals and show how properties of these systems can be extracted by means of non-Markovianity measures.
\end{abstract}

\section{Introduction}
The theory of open quantum systems deals with  the interaction between a small quantum system and a complex environment with many degrees of freedom. Conventionally, the main goal of theoretical investigations on open quantum systems consists of the development of an efficient mathematical description, and corresponding physical analysis, of the open system dynamics by means of phenomenological or microscopic system-environment models. This usually requires the use of various approximation techniques and numerical simulation methods \cite{bp, weiss, zoller}. Very recently a new and interesting perspective has begun to emerge. This revolves around the following question: What can we learn from the dynamics of an open system, playing the role of a quantum probe, about its complex environment?

The possibility of extracting global or local properties of the complex environment by measuring a quantum probe stems from the fact that the probe decoherence and/or dissipation, induced by the interaction with the environment, crucially depends on properties of the latter, such as its spectrum of excitations, its full counting statistics, or its phase in the case of many-body systems undergoing quantum phase transitions.  One can therefore envisage a scenario in which the full control of the quantum probe, under certain conditions, allows for the extraction of information on the complex system, ideally in a non-destructive way. This scenario relies on the following criteria: (i) efficient initial state preparation of the quantum probe; (i) ability to control both the interaction time and the interaction strength between the probe and the complex system; (iii) efficient measurement of the quantum probe dynamics, ideally without affecting the complex system. 

Dramatic developments in the ability to experimentally control both the quantum state of individual particles (atoms or ions) and the coupling to their surroundings allows the identification of physical systems in which the criteria above can be, at least in principle, implemented. Such systems include impurities immersed in ultracold quantum gases and trapped ion crystals. Complex systems such as ultracold atoms and spin chains are generally characterized by highly structured frequency spectra. As a consequence, the decoherent dynamics that they induce in the quantum probe is non-Markovian. The full toolbox of the theory of non-Markovian open quantum systems is needed to describe the dynamics of the quantum probe with sufficient detail. The aim of this review is to introduce these powerful tools and apply them to the scenario on probing complex many-body systems.

During the last five years a number of characterizations and measures of non-Markovianity have been proposed in the literature \cite{eisertmeasure, BLP, RHP, BLPlong, fisher, chinesedivisibility, cv, kossakowski, luomeasure, bogna}, each of them addressing a specific aspect of this multi-faceted and rich phenomenon. The quantification of non-Markovianity is justified by the fact that there is increasing evidence of its important role as a resource for quantum technologies \cite{QKDRuggero, entanglementresource, metrology, teleportation, bogna}. Non-Markovian evolution is often characterized by recoherence phenomena and information trapping, thus leading to longer coherence times in comparison to the Markovian case. Non-Markovianity measures quantify the maximum backflow of information from the system to the environment, hence the non-Markovianity of a quantum probe may be used to infer indirectly, and ideally non-destructively, properties of complex environments such as the occurrence of quantum phase transitions. 

This review is structured as follows. In Sec. II we review basic concepts of the theory of open quantum systems such as the Lindblad theorem, divisibility of the dynamical map, and non-Markovianity measures. In Sec. III we see how a great variety of reservoirs, inducing pure dephasing of impurity probes, can be engineered with ultracold gases. In Sec. IV we present two examples of non-Markovianity as a probe, showing how this quantity allows for singling out the occurrence of a quantum phase transition in Ising models, and a structural phase transition in ion crystals. Finally, in Sec. V we present conclusions.

\section{Non-Markovian dynamics}

\subsection{Open quantum systems}

Quantum systems are never isolated from their surroundings and thus the theory of closed quantum systems fails to describe many essential features of quantum dynamics. It is therefore necessary to include the effect of the environment in the dynamical description of the quantum system. However, including the environment in the equation of motion introduces a large, typically infinite, number of degrees of freedom, complicating tremendously the description of the system. Furthermore, one is typically not interested in the dynamics of the environment but rather on its effects on the system. For this reason it it useful to reduce the description of the total closed system to the description of the system of interest only.

In the theory of open quantum systems one considers a total closed system and separates it into a system of interest, $S$, and its environment, $E$ \cite{bp, weiss, zoller}. The Hilbert space of the total system separates to that of the system and the environment, $\mathcal{H}=\mathcal{H}_S\otimes\mathcal{H}_E$. The dynamics of the open system is obtained from the von Neumann equation of the total closed system: tracing over the environmental degrees of freedom yields a reduced equation of motion for the system, i.e., the master equation
\eq
\frac{d\rho_S(t)}{dt}=-i \mbox{tr}_E\{[H(t),\rho_{SE}(t)]\},
\eeq
where
\eq
\rho_S(t)=\mbox{tr}_E\{\rho_{SE}(t)\}
\eeq
is the state of the system of interest and $\rho_{SE}(t)$ the state of the total system, typically assumed to be initially of the factorised form $\rho_{SE}(0)=\rho_S(0)\otimes\rho_E(0)$. The master equation is exact but often many approximations are required to obtain an equation that can be solved even numerically. The solutions of the master equation at different times correspond to a dynamical mapping
\eq
\Phi_{t_0,t}:\;\rho_S(t_0)\mapsto\rho_S(t)=\Phi_{t_0,t}\rho_S(t_0).
\eeq
The mapping describing the evolution of the open system is, in general, no longer unitary, and it does not conserve the purity of states. This means that the mapping is generally not reversible, and furthermore, positivity of the density matrix is not trivially satisfied. Indeed, the approximations often used to find a more manageable form for the master equation can lead to a mapping that violates the physicality of the ensuing state. This can be manifested as a failure of the mapping to preserve the trace of the positivity of the quantum state, or more subtly, as a violation of the complete positivity (CP) of the state.

Complete positivity is a property of a linear mapping $\Phi:\mathbf{C}^{n\times n}\rightarrow\mathbf{C}^{m\times m}$, where $\mathbf{C}^{n\times n}$ is a $C^*$-algebra of complex $n\times n$-matrices. The mapping is said to be positive if $\Phi(A)>0$ for all $A$, $k$-positive if an extension map $\mathbf{I}_k\otimes\Phi:\mathbf{C}^{k\times k}\otimes\mathbf{C}^{n\times n}\rightarrow\mathbf{C}^{k\times k}\otimes\mathbf{C}^{m\times m}$ is positive and completely positive if it is $k$-positive for all $k$. Intuitively the concept of CP means that the dynamical mapping can be extended to any larger subspace and the mapping still remains positive. It it important to note that a CP mapping is always positive, but the converse is not true in general. 

\subsection{Lindbad theorem}

An important class of open system dynamics for which the physicality (trace preservation and complete positivity) of the evolving state is always guaranteed is given by the Lindblad theorem \cite{linblad, gks}. The dynamical maps forming this class have two interesting properties: they are \emph{time-homogenous}, 
\eq
\Phi_{t_0,t}=\Phi_{t-t_0,0}\equiv \Phi_\tau,
\eeq
where $\tau=t-t_0$, and they obey the \emph{semi-group property},
\eq
\Phi_{t+s}=\Phi_t\Phi_s.
\eeq
Under these two conditions the general form of the dynamical mapping is 
\eq
\Phi_t=e^{\mathcal{L}t},
\eeq
where $\mathcal{L}$ is a bounded generator, and a formal derivation yields the general equation of motion
\eq
\frac{d\rho_S(t)}{dt}=\mathcal{L}\rho_S(t).
\eeq
The theorems of Lindblad \cite{linblad} and Gorini, Kossakowski and Sudarshan \cite{gks} give the most general form of the generator of the dynamical semi-group $\mathcal{L}$. For convenience we cite their common result as the Lindblad theorem and the general form of the operator as the \emph{Lindblad form}, which is explicitly expressed as
\eq \label{lindbladform}
\mathcal{L}\rho_S(t)=-i[H_S,\rho_S(t)]+\sum_k\gamma_k\left[ A_k\rho_S(t)A_k^\dg-\frac{1}{2}\left\{ A_k^\dg A_k,\rho_S(t)\right\}\right],
\eeq
where $\{A,B\}=AB+BA$ is the anti-commutator of two operators, $H_S$ and $A_k$ are bounded operators and $\gamma_k\geq0$ are positive constants. The Lindblad theorem, shown here without the proof, asserts that the following two statements are equivalent:
\begin{enumerate}[(i)]
  \item $\{e^{\mathcal{L}t}\;|\;t\geq0\}$ is a semi-group and for each $t$ the mapping $e^{\mathcal{L}t}:\rho_S(0)\mapsto\rho_S(t)=e^{\mathcal{L}t}\rho_S(0)$ is CP.
  \item $\mathcal{L}\rho_S(t)$ is in the Lindblad form (\ref{lindbladform}).
\end{enumerate}

The implications of the Lindblad theorem are immediate: if a completely positive dynamical mapping $\Phi_t$ fulfills the semi-group condition then we can immediately state the general form of the generator of the semi-group. Namely, the generator is in the Lindblad form. Conversely, if we have an equation of motion in the Lindblad form we are assured that the evolution of the density matrix is always physical as a direct consequence of complete positivity of the dynamical map.

From a physical perspective, how can one interpret a dynamical process described by the Lindblad form? As a first step consider the extreme case of $\gamma_k=0\;\forall k$. In this case the master equation reduces to the von Neumann equation, describing unitary system dynamics of a closed system. The effect of the environment is therefore explicitly contained in the terms of the Lindblad form with $\gamma_k>0$. Each pair $(\gamma_k,A_k)$ corresponds to a different decay channel, e.g., $A_k=\sigma_-$ describing the spontaneous emission of a two-level atom. These decay channels arise naturally in many physical scenarios. Employing the commonly used Born (weak coupling) and Markov (negligibly short system-environment correlation time) approximations in a microscopic derivation of the master equation results in a master equation in the Lindblad form (See, e.g. Ref. \cite{bp}).

Moreover, a master equation in the Lindblad form has a physical interpretation that can be understood in terms of so-called quantum jumps, which we briefly recall here \cite{jump, jump2}. Let us consider a quantum system in a pure state $\ket{\phi}$ evolving during a small interval of time $\Delta t$ according to one of the following processes:
\begin{enumerate}[(1)]
  \item A quantum jump determined by an operator $A$:
  \eq
  \ket{\phi(t)}\mapsto \ket{\phi(t+\Delta t)}_{QJ}=\frac{A  \ket{\phi(t)}}{|| A\ket{\phi(t)}||}
  \eeq
  with a probability of $\Delta p= \langle \phi(t)|A^\dg A |\phi(t)\rangle \gamma \Delta t$, where $\gamma$ is a positive constant.
  \item Unitary evolution generated by an effective non-Hermitian Hamiltonian $H_{eff} = H_S - i \gamma A^\dg A$
    \eq
  \ket{\phi(t)}\mapsto \ket{\phi(t+\Delta t)}_{U}=\frac{e^{-i H_{eff}\Delta t}  \ket{\phi(t)}}{|| e^{-i H_{eff}\Delta t}\ket{\phi(t)}||}
  \eeq
  with a probability $1-\Delta p$.
\end{enumerate}
These two processes amount to the following change of the density matrix corresponding to the state vector:
\eq \label{jump}
\ket{\phi(t)}\bra{\phi(t)}\mapsto \Delta p \ket{\phi(t)}_{QJ}\bra{\phi(t)}_{QJ}+(1- \Delta p) \ket{\phi(t)}_{U}\bra{\phi(t)}_{U}
\eeq
Consider a Monte Carlo wave function process to simulate the evolution of the state vector, where at each interval of time a random number $r\in [0, 1]$ is compared to the probability $\Delta p$ to determine which process, (1) or (2), takes place: when $r<\Delta p$ the state vector undergoes a quantum jump and when $r\geq\Delta p$ the state vector evolves deterministically. When this simulation process is performed for a large ensemble of state vectors, we obtain from (\ref{jump}), in the limit $\Delta t\rightarrow 0$, a master equation that is in the Lindblad form:
\eq 
\frac{d\rho(t)}{dt}=-i[H_,\rho(t)]+\gamma\left[ A\rho(t)A^\dg-\frac{1}{2}\left\{ A^\dg A,\rho(t)\right\}\right],
\eeq
where $\rho(t)=\frac{1}{N}\ket{\psi(t)}_i\bra{\psi(t)}_i$ and $N$ is the number of state vectors forming the ensemble. This simple study shows that a master equation in the Lindblad form describes the dynamics of a quantum system whose deterministic evolution is disrupted by quantum jumps. The constant $\gamma$ is interpreted as the decay rate of that particular quantum jump channel. This result can be generalized to include more than one jump channel, giving the full Lindblad master equation (\ref{lindbladform}).

Master equations in the Lindblad form and the dynamical maps with the semi-group property are especially significant when discussing the border between Markovian and non-Markovian dynamical processes. Indeed, the semi-group property means that the map can be divided into infinitely many time-steps, each identical and independent of the past and future steps \cite{wolf}, and therefore the dynamical map has the intuitive interpretation of memoryless dynamics. These memoryless dynamical maps are commonly called \emph{Markovian}. 

Markovian processes successfully describe a plethora of physical processes, particularly in the field of quantum optics, but can fail if applied to more complex system-environment interactions where memory effects become important. This is typically manifested as long system-environment correlation times that lead to the failure of the Markov approximation. In such situations one must resort to non-Markovian dynamical maps, and the rest of the article focuses on dynamical scenarios where the semi-group Lindblad description of the system evolution is not sufficient to describe the dynamics of the system.

\subsection{Non-Markovianity as non-divisibility}
The Lindblad master equation is often considered the prototype of Markovian, memoryless dynamics. This notion is prompted by the semi-group property which implies that future evolution of the state is independent of the past states. Discrepancies arise when addressing the question of what is \emph{not} Markovian and, to date, a myriad of deviating points of view have been advocated. The discussion on systematic definitions of non-Markovianity is fuelled by the wish to quantify the degree of non-Markovianity in quantum processes. 

The first step in this direction was taken by M. M. Wolf \emph{et al.}, who proposed using deviation from the semi-group property as the principal characteristic of non-Markovian dynamical maps \cite{eisertmeasure}. They further constructed a quantitative measure of non-Markovianity as the minimal amount of isotropic noise that has to be added to the dynamics of an open quantum system to make it Markovian. However, this definition of non-Markovian processes is very severe and, in some cases, open to debate. There are dynamical processes that do not satisfy the semigroup property, but behave in a way that one would intuitively call Markovian.

To make this point more transparent, consider the microscopic derivation of the master equation without the Markov approximation. The ensuing master equation has the same structure as the Lindblad master equation, but with \emph{time-dependent coefficients} $\gamma_k=\gamma_k(t)$. As a result, the corresponding dynamical map is no longer time-homogenous and the semi-group property is violated. However, in the case when the decay rates are positive, $\gamma_k(t)\geq0$ for each $k$ and $t\geq t_0$, the dynamical map has the property of being \emph{divisible}:
\eq \label{divisibility}
\Phi_{t_0,t}=\Phi_{t_0,s}\Phi_{s,t}
\eeq
where $t\geq s\geq t_0$. The dynamical map can thus be concatenated into a collection of other dynamical maps, and analogously to the semi-group property, this concatenation has the intuitive interpretation of memoryless dynamics. The quantum jump picture of the dynamics still holds, albeit now with time-dependent jump probabilities.

Instead, if a Lindblad structured master equation with time-dependent rates has at least one decay rate that takes temporarily negative values, the divisibility of the corresponding dynamical map is broken. In this case the intermediate map $\Phi_{t,s}$ in the concatenation (\ref{divisibility}) is no longer completely positive. Moreover, it can be shown that the converse statement also holds, i.e., non-divisibility is always manifested as a decay rate taking temporarily negative values \cite{RHP}. When a decay rate takes negative values the standard quantum jump picture breaks down due to the appearance of negative probabilities, and an extension thereof, the non-Markovian quantum jump method, becomes necessary \cite{nmqj}. According to this description, during a period of the decay rate being negative a previously occurred jump may be reversed. Reversed jumps recreate earlier states, advocating the idea of memory effects. 

Several authors have put forward the idea of constructing a measure of non-Markovianity based on non-divisibility of the dynamical map \cite{RHP, chinesedivisibility, kossakowski, luomeasure}. The degree of non-Markovianity is quantified by the deviation of the intermediate map $\Phi_{t,s}$ from a completely positive map, which may be measured in several different ways. In this article, however, we focus on a subtly different way of characterising and quantifying non-Markovianity in terms of information flowing from the system to the environment, and back to the system.

\subsection{Non-Markovanity as reversed information flow}

Divisible dynamical processes lead to monotonic decay of many important physical quantities. The converse does not necessarily hold but non-monotonic dynamics of these quantities can still be used to witness non-divisibility. However, non-monotonic dynamics is often considered a signature of non-Markovian dynamics and, especially in the case when one may wish to harness the temporary revivals of these quantities in certain protocols, it is useful to \emph{define} non-Markovian phenomena as deviations from monotonic dynamics.

As an illustration of this approach, we consider here the seminal proposal of Breuer, Laine and Piilo (BLP) \cite{BLP, BLPlong}, where the quantity of interest is the distinguishability of two quantum states $\rho_{1,2}(t_0)$ evolving under the dynamical map $\Phi_{t_0,t}$. Note that henceforth we drop the subscript $S$ from the density matrix of the system $\rho_S(t)$ and refer it to just $\rho(t)$ for brevity. The distinguishability is defined using the trace norm, 
\eq 
D(t)=\frac{1}{2}||\rho_1(t)-\rho_2(t)||_1,\quad \rho_{1,2}(t)=\Phi_{t_0,t}\rho_{1,2}(t_0).
\eeq
A divisible dynamical process decreases the distinguishability of the states monotonically which can be interpreted as a continuous flow of information from the system to the environment. BLP define a dynamical process to be non-Markovian if there is a pair of initial states $\rho_{1,2}(t_0)$ such that the distinguishability increases for at least one interval of time $t\in[a,b]$, taking this to mean that information flows back to the system from the environment. Based on this idea, one can define a measure for the degree of non-Markovianity in a dynamical process as
\eq \label{BLP}
\mathcal{N}=\max_{\rho_{1,2}(t_0)}\int_{\sigma>0}ds\,\sigma(s),\quad\text{where}\quad\sigma(t)=\frac{dD(t)}{dt},
\eeq
and $\sigma(t)$ is called the information flux. The optimisation over all pairs of initial states makes the measure complicated to compute, although it has been shown that the two states maximizing the measure are on the boundary of the state space and orthogonal to each other \cite{anttisteffen}. Moreover, for some simple dynamical maps the maximizing pair has been found \cite{maximisingpair, maximisingpair2, jc}.

This measure of non-Markovianity has been used extensively to study non-Markovian phenomena (see, e.g., \cite{carlosPRL, spinchain, pinja4, pinja5, spinboson, carlos, pinja9}), but it is worth noting that other similar measures have also been introduced. A generalization to continuous variable systems exists, where fidelity is used instead of the trace distance to describe distinguishability of two evolving states \cite{cv}. Lu, Wang and Sun measure information flux in terms of the Fisher information \cite{fisher}. In this scenario one tries to estimate a given quantity, typically the phase of a state, after the state has been evolving under the action of a dynamical map. A lower bound on the variance of this estimate is given by the Fisher information, which decays monotonically when the dynamical map is divisible. Again, one may define non-Markovian processes as those that temporarily increase the Fisher information and integrate over all intervals of positive information flux to give a number quantifying the degree of non-Markovianity. 

Rivas, Plenio and Huelga propose studying the dynamics of a bipartite state comprising of the system evolving under the dynamical map of interest and coupled to a stationary ancilla state \cite{RHP}. The bipartite system is initially in a maximally entangled state and when the dynamical map is divisible, entanglement decreases monotonically. Temporary increase of the entanglement can be interpreted as a non-Markovian effect, and an integration over all such time intervals can be used as a measure of non-Markovianity. The system-ancilla correlations can also be measured by mutual information, as in Ref. \cite{luomeasure}.  Finally, in a similar spirit a very recent proposal studies the non-monotonic behaviour of quantum and classical channel capacities, establishing a link between the non-Markovianity of a quantum channel and the maximal amount of quantum or classical information it can transmit \cite{bogna}.

For the remaining part of the article we take the trace distance based definition and measure (\ref{BLP}) to be the figure of merit in our discussion on non-Markovian phenomena, since it is easy to compute and has a clear and intuitive physical interpretation. It is worth pointing out that for most models considered in this article this definition exactly coincides with non-divisibility of the dynamical map.

\subsection{Microscopic origin of non-Markovianity}

With rigorous definitions of non-Markovianity at hand it is possible to get a deeper understanding of the origin of memory effects. For example, for certain classes of physical models, non-Markovianity can be traced back to specific microscopic features of the model. This knowledge is fundamentally important for reservoir engineering techniques, since it reveals exactly the physical parameters that correspond to (non-)Markovian dynamics in specific realizations of open system dynamics. This paves the way to controlling the appearance and degree of non-Markovianity in these processes. In this section we derive systematically a condition of the spectral density function of a purely dephasing model to have non-Markovian dynamics \cite{pinja7}. It is later shown how this condition relates to certain physical parameters in a realization of the purely dephasing model.

Consder a general dephasing model describing the interaction between a qubit and a bosonic reservoir \cite{luczka, PSE, reina}. The Hamiltonian for this model is
 \begin{eqnarray} \label{dephasinghamiltonian}
 H= \omega_0 \sigma_z+ \sum_k  \omega_k a^{\dag}_k a_k + \sum_k  \sigma_z (g_k a_k+ g_k^* a^{\dag}_k), \nonumber
 \end{eqnarray}
where $\omega_0$ is the qubit frequency, $\omega_k$ the frequenciey of the $k$-th reservoir modes, $a_k \;(a_k^{\dag})$ are the annihilation (creation) operators and $g_k$ describes the coupling constant between the $k$-th reservoir mode and the qubit. In the limit of a continuum of modes the coupling constants are replaced by the the reservoir spectral density function
\eq
J(\omega)=\sum_k |g_k|^2 \delta (\omega_k-\omega).
\eeq
This model admits an exact solution (see Refs. \cite{luczka, PSE,reina}) with constant diagonal elements of the qubit and off-diagonal elements decaying as $\rho_{ij} (t) = e^{-\Gamma(t)} \rho_{ij} (0),\, i\neq j $. The dephasing factor describing this decay for an environment in thermal equilibrium at temperature $T$ is 
\begin{eqnarray} \label{factor}
\Gamma(t) &=& 2 \int_0^{\infty} d \omega\, J(\omega)\coth \left[ \hbar \omega/2 k_B T\right] [1- \cos (\omega t)]/\omega^2, 
\end{eqnarray}
corresponding to a master equation of the form
\begin{eqnarray} \label{exact dissipator}
\frac{d\rho(t)}{dt}=\frac{\gamma(t)}{2} \left[\sigma_z \rho(t) \sigma_z - \frac{1}{2}\left\{\sigma_z\sigma_z,\rho(t)\right\}\right].
\end{eqnarray}
Here the time-dependent dephasing rate is 
\eq
\gamma(t)=\frac{d\Gamma(t)}{dt}.
\eeq
It has been shown that for this dynamical process the pair of initial states optimizing the measure of Eq. (\ref{BLP}) is any pair of antipodal states on the equator of the Bloch sphere \cite{maximisingpair, maximisingpair2}. Without a loss of generality we may choose the initial pair to be $\rho_1(0)=\ket{+}\bra{+}$ and $\rho_2(0)=\ket{-}\bra{-}$, where $\ket{\pm}=(\ket{1}\pm\ket{0})/\sqrt{2}$. For this optimal choice the distinguishability (which for qubit states is half the Euclidian distance of the corresponding Bloch vectors) is 
\eq
D_{\text{opt}}(t)=e^{-\Gamma(t)},
\eeq
leading to optimised information flux
\eqa
\sigma_{\text{opt}}(t)&=&\frac{D_{\text{opt}}(t)}{dt}=-\gamma(t)e^{-\Gamma(t)}.
\eeqa
The qubit dynamics is non-Markovian ($\sigma_{\text{opt}}(t)>0$ for some interval of time) if and only if the dephasing rate takes temporarily negative values, $\gamma(t)<0$. During these intervals the divisibility property is violated and information flows from the environment back to the system as manifested in temporary increase in the distinguishability of the initial states. 

With the exact solution (\ref{factor}) in terms of the spectrum $J(\omega)$ at hand we can formulate a condition for the spectral density to induce non-Markovian dynamics of the qubit. A strictly positive dephasing rate corresponds to monotonic dynamics of the dephasing factor $\Gamma(t)$. Since the cosine transform of a convex function increases monotonically, we can deduce that a sufficient condition for Markovianity for this dephasing model is that the quantity
\eq
\xi(\omega,T)\equiv J(\omega)\coth \left[ \hbar \omega/2 k_B T\right]/\omega^2
\eeq
is a convex function. Furthermore, this condition turns out to be also necessary if we specify the study to the family of Ohmic spectra \cite{weiss}
\begin{eqnarray} \label{ohmic}
J(\omega)=  \frac{\omega^s}{\omega_c^{s-1}} e^{-\omega/\omega_c }, 
\end{eqnarray}
where $\omega_c$ is the reservoir cutoff frequency and $s$ is the so-called \emph{Ohmicity parameter}. The Ohmicity parameter determines if the spectrum is sub-Ohmic ($0<s<1$), Ohmic ($s=1$) or super-Ohmic ($s>1$). For the Ohmic class of spectra the dephasing rate can be calculated analytically in the zero-T and high-T limits:
\begin{equation}\label{rate}
\gamma(t) =\left\{\begin{array}{ll}
\omega_c [1+(\omega_c t)^2]^{-s/2} \Xi[s] \sin \left[ s \arctan (\omega_c t)\right],&T=0 \\
2 k_B T [1+(\omega_c t)^2]^{-(s-1)/2} \Xi[s-1] \sin \left[ (s-1) \arctan (\omega_c t)\right],&\text{high-T},
\end{array}\right.
\end{equation}
where $\Xi[x]$ is the Euler gamma function. 

It is easy to check that $\gamma(t)$ takes temporarily negative values, i.e., the dephasing dynamics is non-Markovian, when $s>2$ in the zero-T case, and when $s>3$ in the high-T case. Therefore the emergence of memory effects for the Ohmic class of spectra comes from the interplay between the Ohmicity parameter $s$ and the temperature of the reservoir. There is always a threshold value of the Ohmicity parameter $s_{\text{crit}}$, depending on the temperature $T$, such that when $s\leq s_{\text{crit}}$ the dynamics is Markovian and when $s> s_{\text{crit}}$ it is non-Markovian.

Interestingly, it can be shown for the zero-T and the high-T cases that the quantity $\xi(\omega, T)$ is non-convex exactly when the qubit dynamics is non-Markovian. The intermediate-T case cannot be studied analytically, but numerical studies confirm that transition of the dynamics from Markovian to non-Markovian coincides exactly with the $\xi(\omega, T)$-function turning from convex to non-convex. Therefore it can be concluded that the convexity of $\xi(\omega, T)$ is a necessary and sufficient condition for Markovianity of the purely dephasing qubit model, in the case of the Ohmic class of spectral density functions. For more general classes of spectra the condition is guaranteed to be sufficient, but the necessity has to be studied separately.

Physically the dephasing process can be understood as follows. The action of the qubit on the environment is a state-dependent displacement operation on each mode of the environment. The two states of the qubit excite each mode with opposite phases and this leads to an overlap between the states of the mode in each case. Destructive interference between excitations of a mode at different times leads to recoherences at the frequency of that mode. The balance between these two effects determines whether the dynamics is Markovian or not. In the case of the Ohmic class of spectra, convexity of $\xi(\omega,T)$ not only guarantees that decoherence outweighs recoherence, but it is required. This highlights the key role of the low frequency part of the spectrum in the occurrence of information backflow for dephasing processes.

This section showed the existence of an explicit connection between the emergence of memory effects in the purely dephasing dynamics of a qubit, and the spectral density function characterising the dynamical process. More generally non-Markovianity is often associated with structured reservoirs, i.e., models where the spectral density function $J(\omega)$ characterising the system-environment coupling varies appreciably with frequency. For the Jaynes-Cummings model, for example, the more structure the reservoir has (the narrower in the Lorenzian spectrum of the cavity), the more pronounced are the memory effects \cite{jc}. However, the connection between spectral densities and non-Markovianity can be much more subtle, as is demonstrated in the case of purely dissipative qubit dynamics, where the low frequency part of the spectrum determines the (non-)Markovian character of the system evolution. 

\section{Simulating non-Markovian processes with ultracold gases}

The dephasing model of the previous section was introduced as a rather abstract model. It was shown that the transition of this model from a Markovian to a non-Markovian process can be controlled by choosing the spectral density function appropriately. For the class of Ohmic spectra, for example, it was shown that Ohmic and sub-Ohmic spectra will always result in Markovian dynamics. Non-Markovianity is only observed for super-Ohmic spectra with Ohmicity parameter $s$ exceeding the critical value $s_{\text{crit}}=2$ for a zero-T reservoir. If thermal effects are taken into account even higher values of the Ohmicity parameter are needed to induce non-Markovianity.

It is natural to ask if there exists a physical scenario realizing this model, and whether such a precise control over the Ohmicity parameter can be realistically achieved. In this section we show that the model of the previous section can indeed be simulated physically using a set-up of ultracold quantum gases \cite{gabriele}. The model is experimentally realistic and, moreover, it can be demonstrated that the effective Ohmicity parameter in this ultracold realisation can be controlled to a very high degree in an experimentally feasible way.

In this section we introduce the ultracold realization of the purely dephasing qubit model \cite{pinja4, pinja6, pinja8}. We unveil the mapping between the model parameters and the physical quantities characterizing the ultracold gases. Specifically we focus on the physical parameters that define the effective Ohmicity parameter, and show how appropriate values of these parameters allow a transition from Markovian to non-Markovian qubit dynamics. This demonstrates how the ultracold gases simulate a fundamental non-Markovian dynamical process.


\begin{figure}
\begin{center}
\label{fig1}
\includegraphics[width=0.4\textwidth]{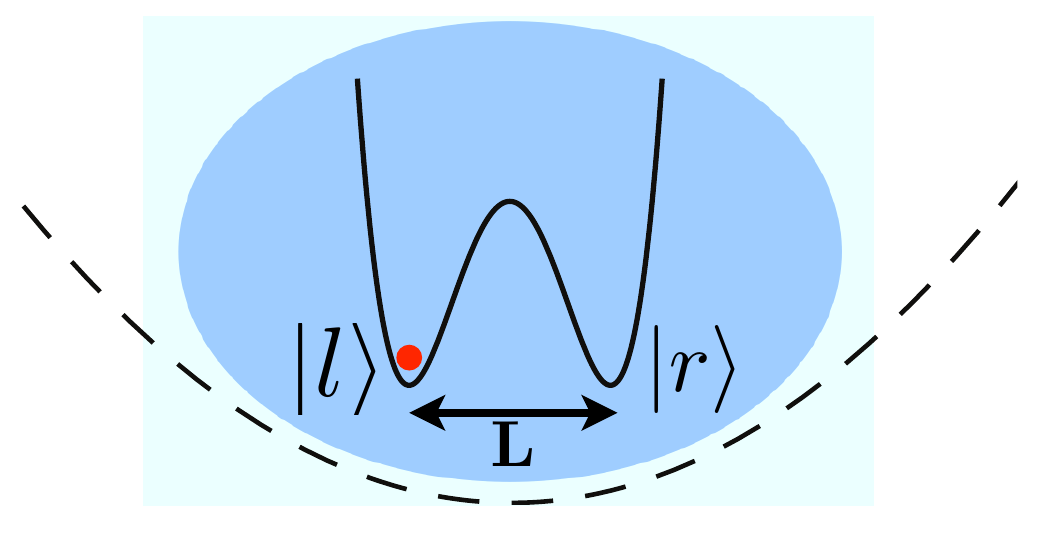}
\end{center}
\caption{Impurity atom (red dot) trapped in a double well potential (solid line), surrounded by a Bose-Einstein condensed ultracold gas of a different species (blue area) trapped in a shallow potential (dashed line).}
\end{figure}

\subsection{The model}
Consider two ultracold gases of different species A and B trapped in optical potentials $V_A(\mathbf{x})$ and $V_B(\mathbf{x})$, respectively. Potential $V_A(\mathbf{x})$ has a very specific form: it consists of a row of deep double well potentials and the ultracold gas of species A is assumed to be so scarce that only a single atom occupies each double well (Fig. \ref{fig1}). A single atom in the double well forms a qubit system with the two qubit states represented by the occupation of the atom in the left or the right well. Potential $V_B(\mathbf{x})$, instead, forms a shallow trap for species B, which is then cooled to a Bose-Einstein condensed (BEC) state. The BEC acts as an environment for the double well qubit, or conversely, the atoms trapped in the double well are impurities in the BEC.  The two different species have a density-density interaction and the Hamiltonian for this system is a sum of the Hamiltonians describing the impurity atoms, the background gas and their interaction,
\eqa \label{ultracoldH}
&&H_A=\int d\xvec\; \Psi^\dg(\xvec)\left[\frac{\mathbf{p}_A^2}{2m_A}+V_A(\xvec)\right]\Psi(\xvec),\nn\\
&&H_B=\int d\xvec\; \Phi^\dg(\xvec)\left[\frac{\mathbf{p}_B^2}{2m_B}+V_B(\xvec)+\frac{g_B}{2}\Phi^\dg(\xvec)\Phi(\xvec)\right]\Phi(\xvec),\nn\\
&&H_{AB}=\frac{g_{AB}}{2}\int d\xvec\; \Phi^\dg(\xvec)\Psi^\dg(\xvec)\Psi(\xvec)\Phi(\xvec),
\eeqa
respectively. Here $m_A$, $\Psi(\xvec)$ and $V_A(\xvec)$ are the mass, field operator and the trapping potential of the impurity atom, $m_B$, $\Phi(\xvec)$, $g_B=4\pi\hbar^2 a_B/m_B$ and $V_B(\xvec)$ are the mass, field operator, coupling constant and trapping potential of a background gas atom and $a_B$ is the scattering length of the boson-boson collisions. Finally, $g_{AB}=4\pi\hbar^2 a_{AB}/m_{AB}$ is the coupling constant of the impurity-boson interaction where $m_{AB}=m_A m_B/(m_A+m_B)$ is the effective mass.

After a series of approximations the total Hamiltonian can be shown to take a form resembling that of Eq. \eqref{dephasinghamiltonian}. The derivation is not shown here in detail. Instead we only sketch the important steps and refer the interested reader to Ref. \cite{gabriele}. The impurity field operator is expanded in terms of Wannier functions $\{\phi_\kvec\}$ localised in the the two wells of the double well potential. Assuming sufficiently deep wells both hopping and tunnelling are suppressed, and the Wannier functions take a Gaussian form. We also assume a weakly interacting background so that it can be treated in the usual Bogoliubov approximation, neglect all terms that are quadratic in the creation and annihilation operators of the Bogoliubov modes and assume that the background gas is homogenous. The interaction Hamiltonian determining the qubit dynamics, up to a possible phase, is then
\eqa
H_{AB}&=&\frac{g_{AB}\sqrt{n_0}}{\Omega}\sum_{\kvec,\;p=L,R}\;\hat{n}_p\, \hat{c}_k\sqrt{\frac{\epsilon_\kvec}{E_\kvec}}\int d\xvec\; |\phi(\xvec_p)|^2e^{i \kvec\cdot\xvec}+H.c.
\eeqa
where $n_0$ is the condensate density, $\Omega$ is the quantization volume, $E_\kvec=\sqrt{\epsilon_\kvec(\epsilon_\kvec+2n_0g_B)}$ is the Bogoliubov dispersion relation, $\epsilon_\kvec=\hbar^2k^2/(2 m_B)$ is the dispersion relation of a non-interacting gas with $k=|\kvec|$, and $\hat{c}_k$ is the Bogoliubov excitation operator. Operator $\hat{n}$ is the number operator of the impurities. When there is exactly one impurity atom in the double well system $\hat{n}_R=\frac{1}{2}(1+\sz)$ and $\hat{n}_L=\frac{1}{2}(1-\sz)$, where $\sz=\ket{l}\bra{l}-\ket{r}\bra{r}$. The two wells are spatially separated by distance $\bf{L}$ so that $\mathbf{x}_R=\mathbf{x}_L-\mathbf{L}$.

The dynamics of a single qubit can be solved analytically \cite{gabriele}. Exactly as in the decoherence model of the previous section, the qubit dephases with $\rho_{ij} (t) = e^{-\Gamma(t)} \rho_{ij} (0),\, i\neq j $. The decoherence factor specific to this set-up is
\eq \label{decoh}
	\Gamma(t) = 8 g_{AB}^2 n_D \sum_{\mathbf k} (|u_k| - |v_k|)^2 e^{-k^2 \sigma^2 /2} \frac{\sin^2 E_k t/2\hbar}{E_k^2} \coth \frac{\beta E_k}{2} \sin^2 ({\mathbf k} \cdot {\mathbf L}),
\eeq
where $\sigma$ is the ground state parameter of the double well, $|u_k|$ and $|v_k|$ are the $k$-th Bogoliubov modes and $\beta = 1/k_B T$. Note that the decoherence factor is explicitly dependent on the dimension of the BEC. This can be altered experimentally by changing the relative strengths of the optical potential $V_B(\mathbf{x})$ in the three axial directions. Stronger confinement in one axial direction creates a flat pancake BEC with effective dimension two. Similarly strong confinement in two axial directions creates an effectively 1-dimensional, cigar shaped BEC. 

Another experimentally feasible way of regulating the qubit dynamics by changing the properties of the environment is provided by the Feshbach resonances. This technique can be used to control the scattering length $a_B$ of the background gas particles very precisely and enables an extrapolation from a free background gas to an interacting background gas. The latter regime is especially attractive from a fundamental perspective. Most studies on open quantum systems focus on non-interacting environments and it is interesting to see the effect of an interacting environment on the emergence of non-Markovianity in the qubit dynamics.

Motivated by real experimental scenarios we consider a $^{87}$Rb-condensate of density $n_3=n_0=10^{20}$m$^{-3}$ and $^{23}$Na impurity atoms trapped in an optical lattice with lattice wavelength $\lambda=600$nm and trap parameter $\sigma=45$nm. The well separation is chosen to be $L=\lambda/8$. The impurity-boson coupling is fixed by setting the corresponding scattering length to $a_{AB}=55\,a_0$, where $a_0$ is the Bohr radius. In the case of a 3D environment the boson-boson coupling frequency is $g_B^{3D}=4\pi\hbar^2a_{B}/m_B$ but now we assume that the s-wave scattering length of the background gas can be tuned from its natural value $a_B=a_{Rb}\approx99a_0$ via Feshbach resonances. We explore a range of values of $a_B$ consistent with the assumption of dilute gas and with the regime of weakly interacting gases. The latter is a stronger condition, requiring $\sqrt{a_B^3 n_0}\ll1$. As a consequence, we can tune the scattering length up to a maximum value given by $a_B\approx3\,a_{Rb}$. The boson-boson coupling frequency is slightly modified for lower dimensions. In the quasi-2D case the scattering length is still much smaller than the axial length of the condensate, $a_B\ll a_z$, and the coupling term is modified to $g^{2D}_B=\sqrt{8 \pi}\hbar^2 a_B/(m_B a_z)$ with 2D condensate density $n_2 = \sqrt{\pi} n_0 a_z$ \cite{stringari}. Within the limits of a dilute gas we can increase the scattering length up to $a_B\approx2\,a_{Rb}.$ The potential $V_B(\mathbf{x})$ can be also modified to create a cigar-shaped quasi-1D background gas with transversal width $a_{\bot}$. The consequent coupling is $g^{1D}_B =2\hbar^2a_B/(m_Ba_{\bot}^2)$ and the 1D density is $n_1=n_0 \pi a_{\bot}^2$, again provided that gas is weakly enough confined, $a_B\ll a_{\bot}$ \cite{bloch review}. In the quasi-1D regime diluteness of the gas allows at most $a_B < a_{Rb}$.

\begin{figure}[h]
\begin{center} \label{fig2}
\includegraphics[width=0.9\textwidth]{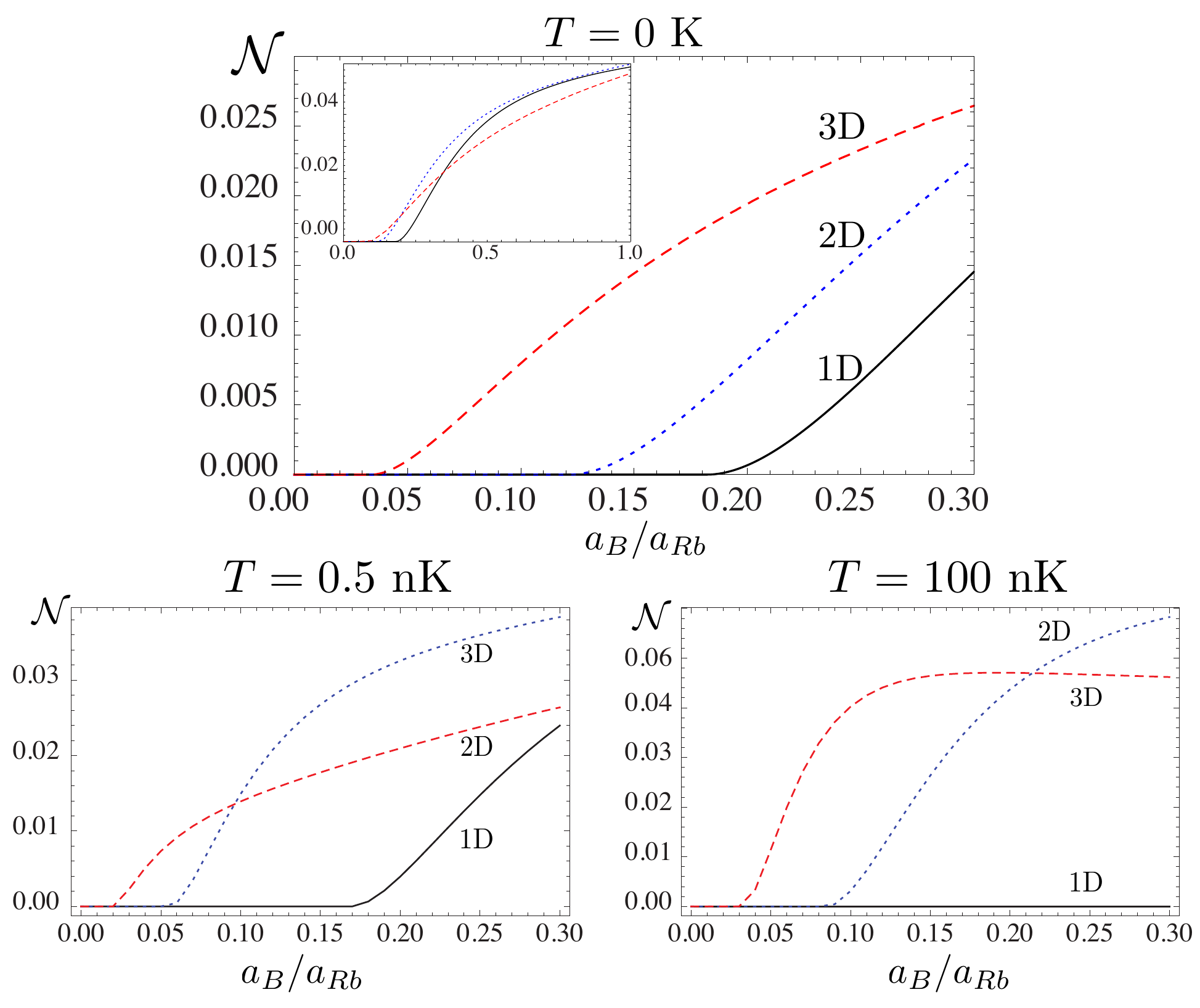}
\caption{Top figure: Non-Markovianity measure $\mathcal{N}$ as a function of the scattering length of the background gas $a_B$ when the background gas is three dimensional (red dashed line), quasi-two dimensional (blue dotted line) and quasi-one dimensional (black solid line) and $T=0$K. The inset shows a longer range of the scattering length $a_B/R_{\text{Rb}}$. Bottom figures: The same for $T = 0.5$nK and $T = 100$nK}
\end{center}
\end{figure}

\subsection{Crossover from Markovian to non-Markovian}

For a realistic set of parameters describing the ultracold gases there is at most only a single period when the decay rate $\gamma(t)=\Gamma'(t)$ is negative, signifying non-Markovianity in the sense of both non-divisibility and information backflow. This enables the use of a modified non-Markovianity measure
\begin{equation}
\label{ultracoldBLP}
\mathcal{N}=\frac{e^{-\Gamma(b)}-e^{-\Gamma(a)}}{e^{-\Gamma(0)}-e^{-\Gamma(a)}},\quad t\in[a,b]\Leftrightarrow\Gamma'(t)>0, 
\end{equation}
which captures the ratio of information returning to the system during interval $a\leq t\leq b$ to the information lost from the system to the environment in the previous interval $0\leq t\leq a$. Unlike the original measure of Eq. (\ref{BLP}), the modified quantifier \eqref{ultracoldBLP} is bounded between zero (system only leaks information) and one (system regains all previously lost information) and is therefore more meaningful as a number. The measure is now a function of the decoherence factor of Eq. (\ref{decoh}).

The non-Markovianity measure for the parameter values elaborated above is shown in Fig. \ref{fig2} in the case of a zero-T reservoir. In all three dimensions the dynamics of the impurity qubit is Markovian for a free or a very weakly interacting background gas, and non-Markovian for a sufficiently large boson-boson interaction of the background gas. The specific critical value of the boson-boson scattering length signifying the crossover from Markovian ($\mathcal{N}=0$, $a_B\leq a_B^{\text{crit}}$) to non-Markovian ($\mathcal{N}>0$, $a_B\geq a_B^{\text{crit}}$) dynamics depends only on the dimensionality of the BEC: $a_B^{\text{crit, 3D}}\approx 0.034\, a_{Rb}<a_B^{\text{crit, 2D}}\approx0.122a_{Rb}<a_B^{\text{crit, 1D}}\approx0.183a_{Rb}$. This strongly indicates that the effective Ohmicity parameter for the ultracold realization of the dephasing qubit dynamics is determined by the scattering length of the background bosons $a_B$ and the dimensionality $D$ of the BEC. Both quantities can be controlled in the laboratory, allowing very precise reservoir engineering and control over the degree of non-Markovianity in this process. 

This analysis can be extended to a finite-T reservoirs, showing that the qubit model here considered is robust enough against thermal fluctuations to retain the crossover from Markovian to non-Markovian dynamics for experimentally realistic temperatures. Only at higher temperatures of about $T=100$ nK the thermal fluctuations start washing out non-Markovian effects, turning the qubit dynamics Markovian. Resilience against thermal fluctuations and the ability to create non-Markovian dynamics with a suitable dimension of the BEC combined with a strong enough boson-boson interaction is a feature characteristic to the double well qubit architecture, as can be shown by comparing the double well model to an atomic quantum dot model.

\subsection{Comparison to AQD}

\begin{figure}[h]
\begin{center} \label{fig3}
\includegraphics[width=0.8\textwidth]{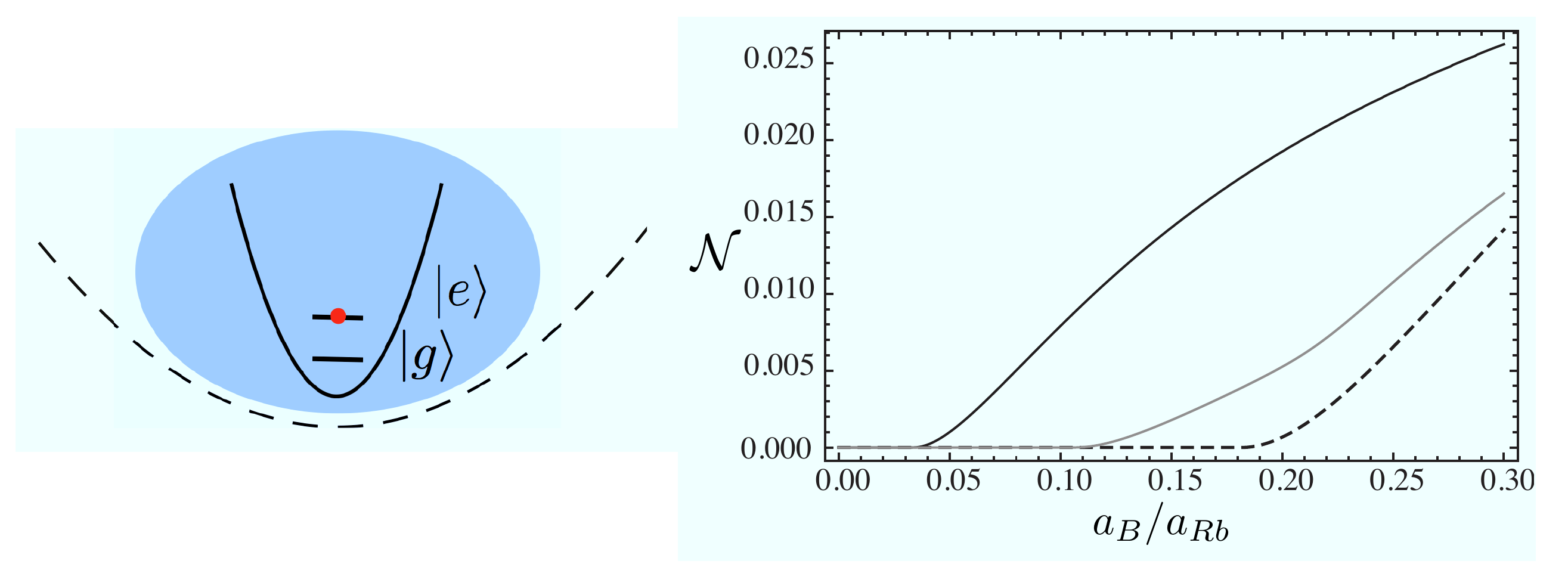}
\caption{Model of an atomic quantum dot and its non-Markovianity measure as a function of the the background scattering length $a_B/a_{\text{Rb}}$ for a zero-T reservoir (solid black line) and a $T = 10 nK $reservoir (solid gray line). Dashed black line shows for comparison the non-Markovianity of the double well qubit in a zero-T reservoir.}
\end{center}
\end{figure}

\begin{figure}[h]
\begin{center} \label{fig4}
\includegraphics[width=\textwidth]{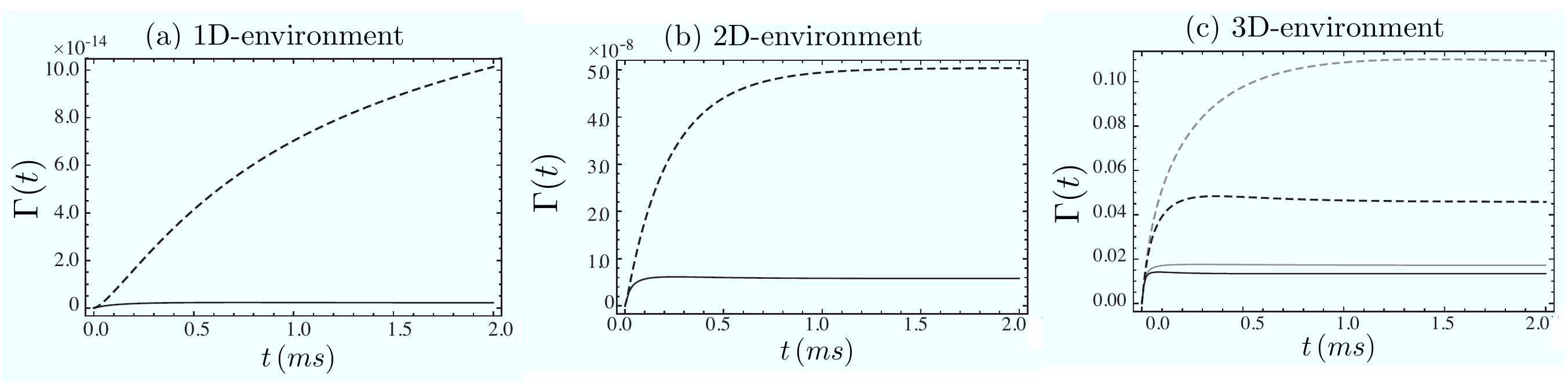}
\caption{The decoherence function $\Gamma(t)$ for the double well qubit (solid lines) and the atomic quantum dot model (dashed lines) for (a) one-dimensional, (b) two-dimensional and (c) three-dimensional BEC environments and for background scattering length values $a_B = 0.25a_{\text{Rb}}$ (black lines) and $a_B =_{\text{Rb}}$ (gray lines).}
\end{center}
\end{figure}

The double well architecture is only one way to realize the pure dephasing model using ultracold gases. A model proposed in Refs. \cite{aqd, aqd2} replaces the double well qubit with a single impurity atom trapped in a deep harmonic potential and takes two internal states of the atom as the qubit states (See Fig. \ref{fig3}). Otherwise the set-up is identical to the model already presented. This model is referred to as the atomic quantum dot and its total Hamiltonian is also given by Eq. \eqref{ultracoldH}.  With the same assumptions as approximations as before the impurity-environment interaction Hamiltonian and the corresponding decoherence factor for the atomic quantum dot model are found to be (see Refs. \cite{aqd, aqd2} for a detailed derivation)
\eqa
H_{AB}&=&\frac{g_{AB}\sqrt{n_0}}{\Omega}\sum_\kvec\;\hat{n}\,\hat{c}_k\sqrt{\frac{\epsilon_\kvec}{E_\kvec}}\int\; d\xvec|\phi(\xvec)|^2e^{i \kvec\cdot\xvec}+H.c.,\nn\\
\Gamma(t)&=&\frac{g_{AB}^2 n_0}{\Omega}\sum_\kvec e^{-k^2\sigma^2/2}\frac{\epsilon_\kvec}{E_\kvec}\frac{\sin^2(\frac{E_\kvec t}{2\hbar})}{E_\kvec^2}.
\eeqa
The similarity between these expressions and those of the double well qubit model is significant. However note that the latter depends on the spatial separation between the two wells of the double well. It turns out that this spatial separation has a crucial impact on the non-Markovian dynamics of the dephasing model.

Figure \ref{fig4} shows the dynamics of the decoherence factor for the two different qubit models. In the cases of two and three dimensional environments the decoherence factors evolve in a similar way, but for the quasi-1D BEC there is a striking dissimilarity in the dephasing process; the double well qubit is almost unaffected by environmental noise (decoherence factors converges quickly to a very small value), while the atomic quantum dot dephases completely (decoherence factor grows without bound), tending to a maximally mixed state. Moreover, the non-Markovian properties of the two models, manifested as non-monotonic dynamics of the decoherence factor, are very different. Dynamics of the atomic quantum dot is Markovian in the quasi-1D and quasi-2D cases and only when the environment is a three dimensional BEC we are able to find a critical value of the scattering length such that there is a crossover from Markovian to non-Markovian dynamics. This is in stark contrast with the double well model which has the crossover in all three dimensions. Moreover, non-Markovian effects in the dynamics of the atomic quantum dot model are not robust against thermal noise. As shown in Fig. \ref{fig3}, a temperature which slightly reduces the amount of non-Markovianity in the dynamics of the double well model washes out all memory effects for the atomic quantum dot.

\subsection{Spectral density function}

\begin{figure}[h]
  \includegraphics[width=0.9\textwidth]{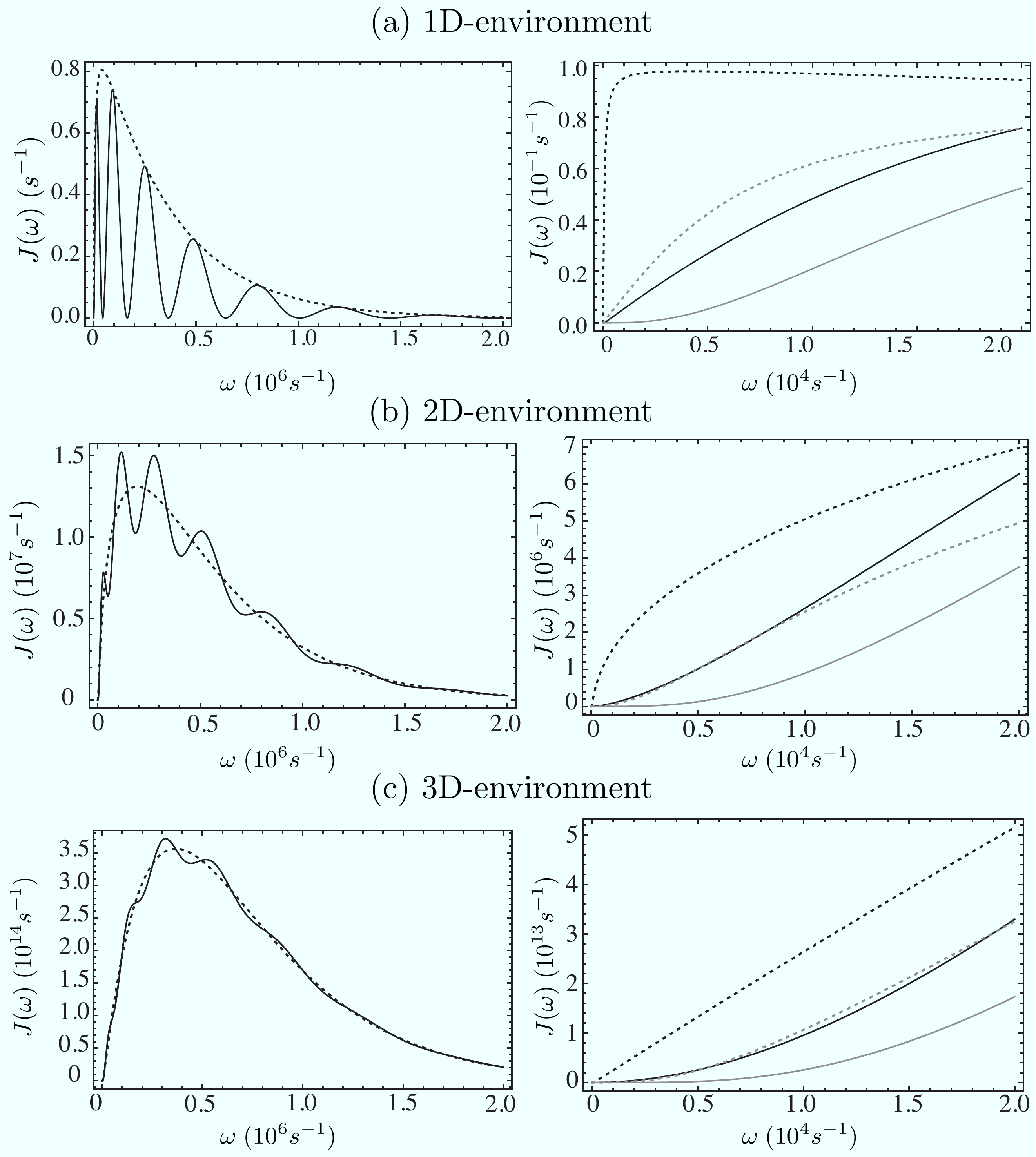}
  \caption{Spectral density functions $J(\omega)$ for the double well model (solid lines in all figures) and the atomic quantum dot model (dashed lines in all figures) in a (a) one-dimensional, (b) two-dimensional and (c) three-dimensional environment. Left hand side figures show the full spectrum, and the figures on the right show the low-frequency contribution. In the latter we show the spectrum for a weakly interacting background gas with $a_B=10^{-3}a_{Rb}$ (black lines) and for a BEC with $a_B=a_{Rb}$ (gray lines).}
  \label{fig5}
\end{figure}

The crossover from Markovian to non-Markovian dynamics for pure dephasing dynamics can be traced back to the behaviour the spectral density function, as discussed in Sec. 2. 5. Indeed, the differences between the to different physical models of qubit dephasing presented here become transparent when we analyze their spectral density functions. Moreover, a comparison of the two spectra contributes in an interesting way to the discussion on the connection between structured spectra and non-Markovian dynamical processes. The differences between the spectral density functions of the double well qubit model and the atomic quantum dot model underline the immense importance of the of the low-frequency part of the spectral density function for dephasing processes.

The form of the spectral density functions for the physical realizations of the dephasing model are generally more complicated than the phenomenological spectrum \eqref{ohmic} of the general dephasing model. The low frequency part of each spectra, however, is well approximated by an Ohmic type dependence on the frequency, 
\eq
J(\omega)\propto \omega^{s_{\text{eff}}}.
\eeq
The \emph{effective} Ohmicity parameter $s_{\text{eff}}$ depends on the model parameters and turns out to be especially sensitive to the dimensionality of the BEC environment and the boson-boson scattering length $a_B$; increasing either of these experimentally controllable parameters increases the value of  $s_{\text{eff}}$ (See Fig. \ref{fig5}). Crucially, with suitable choice of parameters the value of  $s_{\text{eff}}$ is sufficiently large to induce non-Markovian effects, exactly as in the general dephasing model. For both qubit models in a zero-T environment the parameters leading to $s_{\text{eff}}$ correspond to the qubit dephasing in a non-Markovian way. When thermal effects are taken into account, a larger threshold value has to be reached. For a given set of parameters the value of $s_{\text{eff}}$ for the double well qubit is always larger than for the atomic quantum model, explaining why the former model is more non-Markovian than the latter.

As conjectured before, it is the low frequency part of the spectral density function that dictates whether the dynamics of the system is Markovian or not. The extent of this statement becomes evident when we look at the whole spectrum in Fig. \ref{fig5}. The spectral density function of the double well qubit model has a very rich structure over the whole range of relevant frequencies, especially in the case of a 1D environment. Contrary to the typical idea that structured spectra are associated to non-Markovian effects, for dephasing dynamics this structure has no effect on the non-Markovianity. If $s_{\text{eff}}<s_{\text{crit}}$, as determined by a small enough scattering length $a_B$, the qubit dynamics is still Markovian despite the rich structure over the whole range of relevant frequencies. This result highlights the importance of the low frequency part of the spectrum in the emergence of non-Markovian dynamics and emphasizes the difference between pure dephasing dynamics and models such as the driven qubit or the Jaynes-Cummings model, where structures spectra can be associated to memory effects.

\section{Non-Markovian quantum probes}

The archetype of open quantum systems, i.e., a small system of interest interacting with a larger environment, lends itself to an alternative viewpoint. If we are interested, instead, in the larger system, can we infer some of its properties by studying the smaller system? In this picture the smaller system acts as a \emph{probe} of the larger system. Evolution of the probe system is dictated by the properties of the environment and the way the two are coupled, and with a suitable set-up a given property of the environment can be exposed by some feature in the evolution of the probe. In a best-case-scenario this property can be studied without significantly affecting the larger system, in the spirit of making a non-destructive measurement  of the environment. In this section we explore this question first in a general setting, and then by focusing on two different physical models.

\subsection{Non-Markovianity and Lochmidt echo}

Consider once again a purely dephasing open system model, where a qubit is coupled to a complex many-body environment via an interaction term $H_{I}$ \cite{pinja5}. Assume an initially factorized composite state 
\eq
\rho_{SE}(0) =|\phi_{S}\rangle \langle \phi_S|\otimes\rho_{E}(0)
\eeq
with a pure initial system $|\phi_{S}\rangle=c_{g}|g\rangle+c_{e}|e\rangle,\;|c_g|^2+|c_e|^2=1$. The evolution of the environment is consequently split into two branches with weights given by the qubit coefficients $c_\alpha$, $\alpha=g,e$ so that in each branch the environment Hamiltonian $H_E$ is replaced by an effective Hamiltonian 
\eq \label{branches}
H_{\alpha} = H_{E} +\bra{\alpha}H_{I}\ket{\alpha}.
\eeq
Here the latter term represents the action of the qubit on the environment described by the interaction Hamiltonian. Tracing over the environmental degrees of freedom we find that the qubit evolves as 
\eq
\rho_{S}(t)= |c_{g}|^2|g\rangle\langle g|+|c_{e}|^{2}|e\rangle\langle e| +c_{g}^{*}c_{e}\,\nu(t)|e\rangle\langle g| + \mbox{H.c.},
\eeq
where $\nu(t)$ is the decoherence factor. When also the initial environmental state is pure, $\rho^{E}(0) =|\Phi\rangle\langle\Phi|$, the decoherence factor is simply the overlap between perturbed environmental states of the two branches
\eq
\nu(t)= \langle\Phi|e^{i  H_{g}t}e^{-i  H_{e}t}|\Phi\rangle.
\eeq 
The square modulus of the decoherence factor 
\eq
|\nu(t)|^{2}=L(t)
\eeq
is a quantity knows as the Loschmidt echo. Loschmidt echo describes the stiffness of a many-body system to external perturbations and it emerges in many interesting fields of study \cite{loschmidtreview}. In the context of chaotic systems, for example, the Loschmidt echo is used to study the sensitivity of dynamics on the initial state. It also characterizes the ability of a system to return to its initial state after imperfect backwards evolution, thus having consequences on the study of time reversal. Indeed, the concept of Loschmidt echo was originally coined after extensive discussions between Loschmidt and Boltzmann on the origin of macroscopic irreversibility. 

In this case we obtain the Loschmidt echo of the many-body environment directly from the qubit evolution. Furthermore, information flow between the system and the environment is determined by the Loschmidt echo. More explicitly, the pair maximizing the measure of non-Markovianity for dephasing noise is again a pair of antipodal states on the equator of the Bloch sphere and for this optimal pair the distinguishability is
\eq
D_{\text{opt}}(t)=\sqrt{L(t)}.
\eeq
The distinguishability is a monotonic function of the Loschmidt echo and hence any non-monotonic behavior of the Loschmidt echo immediately indicates a reversed information flux ($\sigma_{\text{opt}}(t)>0$). This gives a neat expression for the measure of non-Markovianity in terms of the Loschmidt echo:
\begin{equation}
\label{eq:NL} \mathcal{N}=\sum_{n}
\sqrt{L(b_{n})}-\sqrt{L(a_{n})},
\end{equation}
where $[a_{n},b_{n}]$ are the time intervals over which $L'(t)>0$ and $L(a_n)$ and $L(b_n)$ are the local minimum and maximum, respectively, of the Loschmidt echo. The utility of this simple connection between the non-Markovianity measure and the Loschmidt echo becomes transparent when the environment is a critical system.

\subsection{Ising model in a transverse field}

Ising model in a transverse field is a prototype of a quantum critical system \cite{qpt}. This simple and intuitive model comprises of a 1D chain of spins with nearest neighbour interactions, characterized by parameter $J$, and under the influence of a transverse magnetic field with interaction strength $\lambda$. The Hamiltonian for this system is
\begin{equation}
\label{eq:isingtransverse}
H_{\text{Ising}}=-J\sum_{j=1}^N{\sigma^{z}_{j}}{\sigma^{z}_{j+1}}+\lambda{\sigma^{x}_{j}}.
\end{equation}
In the limit of $\lambda\ll J$ the system has a doubly degenerate ferromagnetic ground state, where all the spins point in a single direction, either up or down. In the opposite limit $\lambda\gg J$ the magnetic field dominates the spin-spin interactions and the system has a paramagnetic ground state with all the spins polarized in the direction of the magnetic field. When the two competing effects balance, $\lambda=J$, the transverse Ising model undergoes a second order quantum phase transition.

The phase transition of the transverse Ising model can be probed with a central spin which couples to all the spins in the Ising chain with equal strength $\delta$ \cite{cpsun}. The interaction Hamiltonian is then
\begin{equation}
H_{I}=\delta|e\rangle\langle e|\sum_{j=1}^N\sigma^{x}_j.
\end{equation}
Inserting this interaction into Eq. (\ref{branches}) shows that the central qubit splits the evolution of the transverse Ising model into two branches. In the branch corresponding to the qubit in the ground state the Ising chain evolves according to Hamiltonian \eqref{eq:isingtransverse}, while in the other branch the strength of the magnetic field is replaced by an effective value $\lambda^*=\lambda+\delta$; the presence of the qubit effectively increases the impact of the transverse field by a small quantity $\delta$. Close to the critical point this small perturbation is enough to significantly alter the dynamics of the Ising chain, which is reflected right back to the dynamics of the central probe qubit.

\begin{figure}[h]
\begin{center} \label{fig6}
  \includegraphics[width=0.55\linewidth]{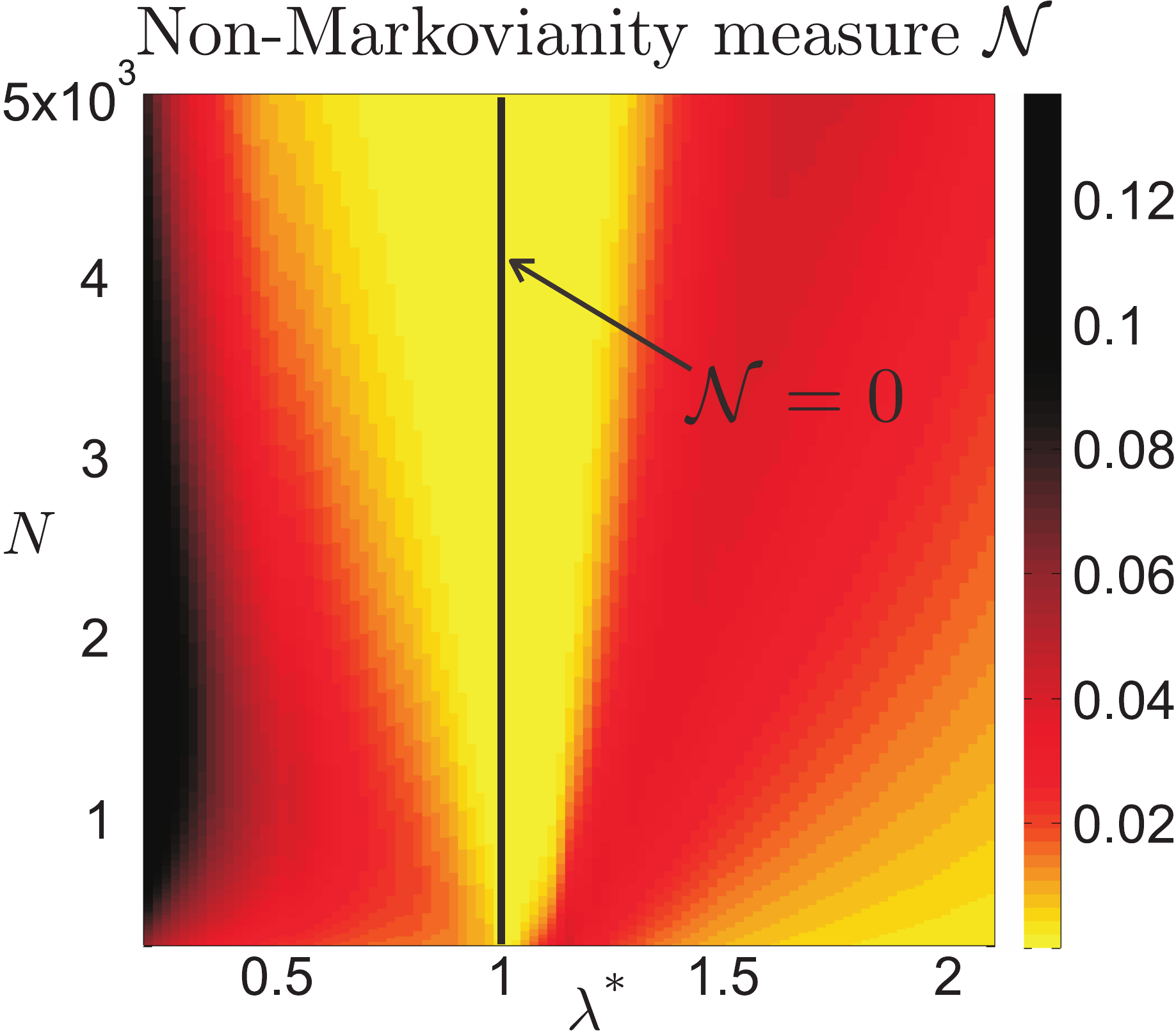}
  \caption{Non-Markovianity measure ${\cal N}$ as a function of the particle number $N$ and the renormalized field $\lambda^*$. }
\end{center}
\end{figure}

It has been shown previously that the decay of the Loschmidt echo is strongly enhanced around the critical point \cite{cpsun}. The hyper-sensitivity of the Loschmidt echo to the critical point translates in a striking way to the non-Markovianity measure (\ref{eq:NL}) of the central qubit. (Again we omit the straightforward but lengthy calculation, preferring to focus on the physical implications of the final result. An interested reader is referred to Ref. \cite{pinja5} for details.) It should be noted that since the Loschmidt echo of the transverse Ising model does not have an analytical expression in the thermodynamic limit, we can only study the non-Markovianity measure for a finite number of environment spins. Consequently the echo will always have revivals arising from finite-size effects and the qubit dynamics is trivially non-Markovian. Fortunately in many cases it is possible to distinguish different physical phenomena happening on different time-scales, e.g., in this case the revivals due to finite-size effects and the revivals due to the structure of the environment. In Fig. \ref{fig6} the time-integration of the non-Markovianity measure has been truncated to a time smaller than the expected reoccurrence time and hence the measure captures revivals due to information flowback. 

A plot of the non-Markovianity measure for various values of the effective field $\lambda^*$ and number of spins N in the Ising chain is shown in Fig. \ref{fig6}. When the magnetic field is tuned in such a way that the Ising model is not at the critical point, the dynamics of the central qubit is always non-Markovian and at least a small amount of information returns temporarily from the environment to the system. The closer the Ising model is to the critical point, the less information is returned to the system. However, it is \emph{only  at the critical point} when the information flow is unidirectional. An especially interesting feature is the independence of the result $\mathcal{N}=0$ on the size of the environment. It can therefore be concluded that the non-Markovianity measure has a strong imprint of the quantum phase transition of the Ising model, even for a finite-sized environment. Indeed, $\mathcal{N}$ has the remarkable property of being able to pinpoint the critical value of the transverse field away from the thermodynamic limit where the quantum phase transition truly takes place. This means that the centrally coupled qubit can be used to probe the quantum phase transition of the Ising model in a transverse field.

\subsection{Coulomb chain}

\begin{figure}[h]
\begin{center} \label{fig7}
  \includegraphics[width=\linewidth]{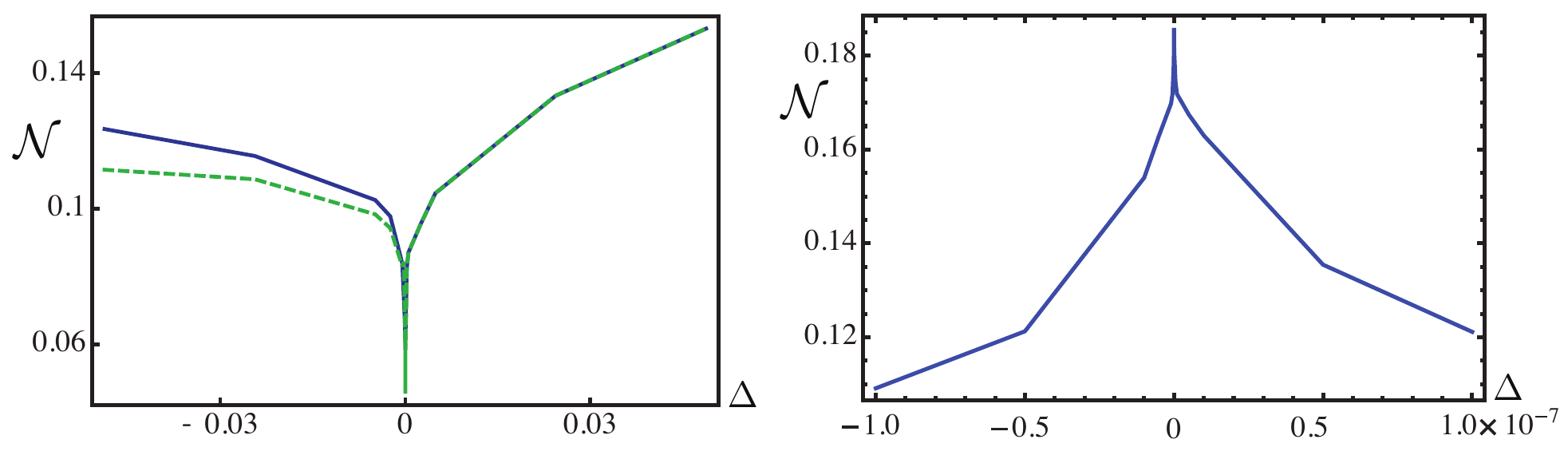}
  \caption{Left: Non-Markovianity measure $\mathcal{N}$ of the Coulomb chain for short time-scale truncation as a function of $\Delta = \nu_t/\nu_c-1$ for $N = 100$ (blue solid line) and $N = 1000$ (green dashed line). $\nu_t$ is the tuning parameter and $\nu_c$ its critical value, i.e., $\Delta$ quantifies the distance from the critical point.
  Right: Non-Markovianity measure $\mathcal{N}$ of the Coulomb chain for long time-scale truncation as a function of $\Delta$  for $N = 300$.}
\end{center}
\end{figure}

The ability of a probe qubit to pinpoint the quantum phase transition of its environment is not unique to the model introduced above. The non-Markovianity measure characterizing the dynamics of a probe qubit can also be used to indicate the phase transition of a many-body system, as sketched in this section.

The system in question is a Coulomb chain of repulsively interacting ions in an anisotropic trap with a large transverse confinement and cooled to very cold temperatures where quantum fluctuations dominate the system properties. With sufficiently strong confinement in the transverse direction the ions form a linear array. If the transverse confinement is reduced or the density of the ions in the chain is increased, the chain undergoes a structural phase transition to a zigzag structure. The precise nature of the structural linear-zigzag phase transition is still unresolved, but there is increasing evidence that it is a quantum phase transition \cite{phi4}, and of the same universality class as the transverse Ising model \cite{gabrielepreprint}.

It has been proposed that this structural phase transition can be observed using Ramsey interferometry on one of the ions in the Coulomb chain \cite{gabrieleramsey}. In this scheme a transverse laser excites the modes of the Coulomb chain and the ion is left to evolve freely with the rest of the chain. After a time $t$ a second laser pulse of the opposite direction is imposed on the probe ion, and the ground state probability $P_g(t)$ is then measured. The ground state probability now depends on the properties of the external degrees of freedom, namely the state of the Coulomb chain, and its evolution is distinctly different for a chain in the linear configuration and for a chain in the zigzag configuration.

The Ramsey interferometry scenario readily admits an open system picture where the two internal degrees of freedom of the ion excited in the Ramsey sequence play the role of a qubit and the excitation modes of the kicked Coulomb chain are the environmental degrees of freedom. The dynamical map describing the dynamics of the qubit system is very complex, including the effect of the two laser pulses of the Ramsey interferometry and the intermediate unitary evolution of the Coulomb chain. 

For this dynamical map the pair of states maximising the non-Markovianity measure has to be resolved numerically since an analytical result for complicated dissipative dynamics is not known. Compelling numerical evidence indicates that the maximizing pair is the same as for the dephasing model, namely an antipodal pair of states on the equator of the Bloch sphere with an equal superposition of the two states $(\ket{e}\pm\ket{g})/\sqrt{2}$. Interestingly, this pair of states is not affected by dissipative dynamics and undergoes pure dephasing dynamics only, making the situation intriguingly similar to the transverse Ising model probed by a centrally coupled qubit. The distinguishability of the optimal pair of states and the consequent non-Markovianity measure depend explicitly on the visibility of the Ramsey interferometry signal, a quantity reminiscent of the Loschmidt echo describing the evolution of the overlap of two environment states. 

The sensitivity of the environment to perturbations close to the critical point, now measured by the visibility, is again manifested in the non-Markovian character of the probe qubit. However, the relationship between the criticality of the Coulomb chain and the non-Markovianity measure is more complex than in the case of the transverse Ising model. In particular, there is now a greater sensitivity to the time truncation in the integration of the BLP measure, as shown in Fig. \ref{fig7}. For short evolution times the non-Markovianity measure has a distinct minimum at the critical point. Unlike in the case of the transverse Ising model the measure does not go to zero at the critical point indicating a total inhibition of information backflow, but nonetheless the reversed flow of information is clearly suppressed.

If the truncation is made after a longer period of time, the dependence of the non-Markovianity measure on the deviation from criticality is rather different: now the measure has a \emph{maximum} at the critical point (See Fig.  \ref{NMcoulomb}). This is explained by a specific feature of the long-time dynamics, namely the system coupling dominantly to only a single "soft" mode of the environment. Coupling to only a single environment mode creates a strong interaction between the mode and the probe qubit leading to considerable bidirectional information exchange and a high value of the non-Markovianity measure. This effect is enhanced the closer the system is to the critical point, explaining why the measure has a maximum at the critical point.

\section{Outlook and conclusions}

In this review we have seen examples of the potential use of non-Markovianity of an immersed quantum probe to extract information on the complex system they are interacting with. We have seen that a two-state impurity immersed in a Bose-Einstein condensate undergoes Markovian or non-Markovian dynamics depending on the strength of the interaction between the atoms of the condensate, and we have explained how, by changing the scattering length of the condensate, one can observe the transition from Markovian to non-Markovian dynamics. As this crossover uniquely depends on the dimensionality of the gas, one can infer the transition from a 3D to a quasi-2D to a quasi-1D condensate simply by measuring the probe. We have then quantitatively linked non-Markovianity to the Loschmidt echo and seen how the former quantity vanishes exactly and only at quantum phase transition for the Ising model in a transverse field. In the Ising model example we consider chains of spins subjected to nearest neighbour interactions. In comparison to the other example considered, the ion crystal, each ion experiences long-range interaction. By changing the trapping potential the ions undergo a structural phase transition from a linear to a zig zag configuration. Also here the non-Markovianity measure is an indicator of the occurrence of a phase transition. 

All the case-studies considered here seem to indicate that it is indeed possible to extract quantum information on a complex system by means of a quantum probe and, in some cases, non-Markovianity is a good quantifier for such a purpose. There are however a number of open questions that wait to be answered. A general theory investigating the advantages and possible limitations of quantum probes is presently missing, as well as a clear understanding of the potential of multiple local quantum probes versus entangled quantum probes. The use of quantum probes may be of particular importance for verifying quantum simulations. By definition, specific-purpose quantum simulators allow for the emulation of quantum systems whose dynamics or general properties cannot be studied with classical computers. Hence, in many cases, one cannot verify the correctness of their result. However, there should be ways of increasing the confidence in the result of the simulation. Most of the existing measurements on ultracold gases either destroy the physical system performing the simulation, e.g., a Bose lattice or, by the very act of observing the many-body system, inevitably project a quantum superposition or entangled state into its corresponding statistical mixture, therefore deleting its quantumness. A way to overcome this problem, in principle, is to avoid a direct measurement of the many-body quantum simulator by detecting its properties in an indirect way by means of a quantum probe. If the conditions  for efficient quantum probing are satisfied, then we will be able to map properties of the quantum simulator onto the quantum probe dynamics and extract this information without destroying the quantum simulator itself. 

While much remains to be done to lay the foundation of a theory of quantum probes for complex systems, we are convinced that such a theory would lead to a paradigm shift in the way we currently think of quantum measurements. 

{\it Acknowledgements.}
We acknowledges financial support from EPSRC (EP/J016349/1), the Finnish Cultural Foundation (Science Workshop on Entanglement), and the Emil Aaltonen foundation (Non-Markovian Quantum Information). We thank Suzanne McEndoo for discussions and for carefully reading the manuscript.

\end{document}